# SODA: Generating SQL for Business Users

Lukas Blunschi
ETH Zurich, Switzerland
lukas.blunschi@inf.ethz.ch

Claudio Jossen
Credit Suisse AG, Switzerland
claudio.r.jossen@credit-suisse.com

Donald Kossmann
ETH Zurich, Switzerland
donald.kossmann@inf.ethz.ch

Magdalini Mori
Credit Suisse AG, Switzerland
magdalini.mori@credit-suisse.com

Kurt Stockinger
Credit Suisse AG, Switzerland
kurt.stockinger@credit-suisse.com

## ABSTRACT

The purpose of data warehouses is to enable business analysts to make better decisions. Over the years the technology has matured and data warehouses have become extremely successful. As a consequence, more and more data has been added to the data warehouses and their schemas have become increasingly complex. These systems still work great in order to generate pre-canned reports. However, with their current complexity, they tend to be a poor match for non tech-savvy business analysts who need answers to ad-hoc queries that were not anticipated.

This paper describes the design, implementation, and experience of the SODA system (Search over DAta Warehouse). SODA bridges the gap between the business needs of analysts and the technical complexity of current data warehouses. SODA enables a Google-like search experience for data warehouses by taking keyword queries of business users and automatically generating executable SQL. The key idea is to use a graph pattern matching algorithm that uses the metadata model of the data warehouse. Our results with real data from a global player in the financial services industry show that SODA produces queries with high precision and recall, and makes it much easier for business users to interactively explore highly-complex data warehouses.

## 1. INTRODUCTION

### 1.1 Problem Statement

Modern data warehouses have grown dramatically in complexity over the last decades. In particular, the schemas of data warehouses have become increasingly complex with hundreds of tables and ten thousands of attributes for many organizations. In part, this growth in complexity has been the result of the large success of data warehousing in many organizations. Data warehouses are used for an increasing number of applications and these applications have evolved over time. Each new application and most evolutionary steps involve extending the schema in order to *fiddle in* the new information requirements of, say, the new application.

A second observation that can be made in modern data warehouses is that there is a growing gap between the high-level (conceptual) view of business users and the low-level (physical) perspective of database administrators. Business users still think of the data in star schemas with fact tables in the center and dimension tables as satellites [13]. Database administrators need to integrate many such star schemas of different kinds of business users with varying information needs into a single physical schema. Their job is to optimize the data warehouse, thereby minimizing cost (i.e., $) and meeting all performance goals (i.e., response time and throughput). At the same time, they must manage the data and the schema.

Given these differences in goals, it is not surprising that the conceptual world of business users and the physical world of database administrators is very different. For instance, database administrators may implement a simple business concept such as *Customer* using many different tables, thereby partitioning the data horizontally and vertically. Furthermore, database administrators may store information from different business entities in a single table if that helps improve performance or manageability. Database administrators may also implement *inheritance* and *generalization* in different ways, depending on the query workload that they anticipate. As an extreme example, database administrators may use cryptic naming schemes for table and column names, thereby helping them with certain administration tasks. What makes matters worse is that the schemas of data warehouses have already evolved for several decades and different conventions and optimizations have been applied in each generation.

In regular, every-day operations, this gap does not become apparent. The information needs of business users are typically fulfilled with the help of pre-defined reports using pre-canned queries. These pre-canned queries specify exactly how to reconstruct the business concepts (e.g., revenue of a customer) from the physical database schema. While these reports work well for periodically recurring information needs of business users, the gap between the business and IT world becomes problematic if business users want to ask ad-hoc queries or if new reports need to be generated for an optimized business processes or to launch a new product. In such an event, business users and database administrators must work together and it often takes days or weeks before both groups of users have found a way to implement such a new report if all the information is already in the data warehouse.

### 1.2 SODA Overview

In order to support a more agile usage of a data warehouse, new search tools are required. Ideally, a business user asks a query using operators and the business concepts of her world and the search tool automatically translates these concepts into SQL queries that are executable in the current version of the data warehouse. Typical queries might be: Show me all my *wealthy customers* who live in Zurich. Who are my *top ten customers* in terms of revenue? In such queries, *wealthy customers* is a business concept that is defined by, say, the *salary* of a customer. *top ten* is an operator and applied to *customer* it asks for the customers with the highest trading volume.

SODA addresses this need of business users by allowing them to pose queries in an intuitive, high-level language based on keywords, operators and values. SODA translates these queries into a set of alternative SQL queries, ranks those queries, and (partially) executes the Top 10 in order to generate result snippets (up to twenty tuples) for each of these queries. Just as in a Web search with Google or Bing, the user has now the choice to select one of those queries of the first result page, ask for the next set of candidate queries (i.e., the next result page), or refine the original query.

Translating keywords into SQL has been studied before in related work such as BANKS [3], DISCOVER [10], DBExplorer [1], SQAK [23] or Keymantic [2]. Like most of these systems, SODA indexes the base data and finds join paths using key/foreign key relationships of the database schema. The key innovation of SODA is its flexible way of making use of metadata that goes way beyond looking at key/foreign key relationships or lookups on column names and table names. SODA allows to define metadata patterns that specify how the database schema implements the conceptual model that the business user might have in mind. For instance, at Credit Suisse customer information is spread across several tables; different kinds of customers (e.g., organizations, wealthy customers in private equity, customers with special compliance constraints and risk profiles) are implemented in different ways. The metadata allows to bridge the gap between the low-level SQL implementation and the concepts typically used by business users and allows to generate the right SQL for a complex query.

To enable SODA (and other related tools), Credit Suisse has invested heavily in building a so-called *metadata warehouse* [11]. Such a metadata warehouse stores all available metadata. [11] shows, for instance, how provenance information of the metadata warehouse can be used in order to find out which applications are affected by a change in a specific data source. As will be shown, SODA exploits the definition of business terms (e.g., *wealthy customer*), homonyms and synonyms (e.g., information extracted from DBPedia), and data models at different levels to help business users to ask complex queries to the data warehouse. Given the growing gap between business and IT, we are aware of several other organizations that are also investing into such metadata warehouses. Obviously, these metadata warehouses will have totally different structures and model metadata in different ways. By using patterns, however, SODA is flexible and generic enough to exploit any kind of metadata. Furthermore, SODA can evolve over time thereby adapting to new patterns based on user feedback or to an evolving data warehouse.

### 1.3 Contributions and Overview

In summary, the main contributions of this paper are as follows:

a) This paper shows how SODA can be used to generate SQL queries from a high-level query language (keywords and operators), metadata, and patterns. The key innovation is the use of patterns that help to interpret and exploit a large variety of different kinds of metadata such as homonyms and synonyms (e.g., using DBPedia), domain ontologies (Credit Suisse has its own domain ontology and imports standards from the financial industry), modeling conventions (e.g., inheritance), performance tricks (e.g., partitioning, redundancy), and last but not least base data.

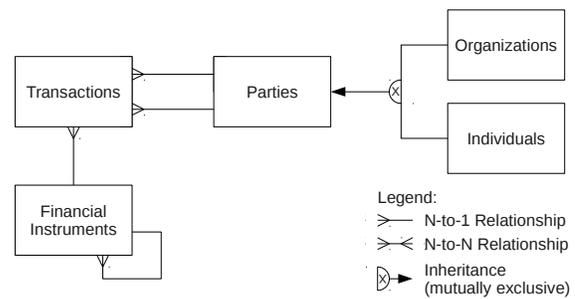

**Figure 1: Sample World: Conceptual Schema.**

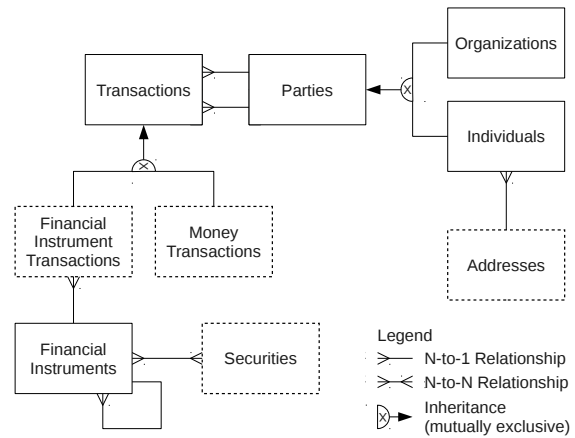

**Figure 2: Sample World: Logical Schema.**

b) This paper presents the results of experiments using a real-life data warehouse with hundreds of tables and thousands of attributes from a major player in the financial industry. The experiments show that indeed complex SQL queries can be generated automatically from high-level specifications and that the gap between the world of business users and the reality of IT can be bridged. The experiments, however, also demonstrate some of the limitations of the approach and that a search tool like SODA needs to evolve based on user feedback and experience.

The remainder of this paper is structured as follows: Section 2 describes the running example used throughout this paper. Section 3 gives an overview of the SODA approach. Section 4 defines the algorithms used in SODA in more detail and presents several example queries. Section 5 discusses the results of experiments that evaluate the quality of the search results produced by SODA. Section 6 compares SODA to related work; in particular, Section 6 shows how existing systems would fare for the queries used in our experiments. Section 7 contains conclusions and possible avenues for future work.

## 2. RUNNING EXAMPLE

This section describes the simplified schema of a mini-bank with customers that buy and sell banking products (so-called financial instruments). We use this example throughout this work to demonstrate that generating SQL queries that meet a business need can be difficult for humans even for a small schema. The example, however, also illustrates that if the metadata is known and the right *patterns* can be extracted from the query, then generating the right SQL is quite doable for a machine.

Typical end user queries that we will analyze throughout the paper are as follows: (1) Find all financial instruments of customers in Zurich. (2) What is the total trading volume over the last months? (3) What is the address of Sara Guttinger?

## 2.1 Example Schema

Figure 1 shows the example schema. It models information about customers (referred to as *Parties* in our data warehouse) and the transactions these customers made; buying and selling on the stock market. Parties can be *individuals* for private banking or corporate customers for investment banking (i.e., *organizations*); both kinds of parties are modeled separately because they are supported by different sets of analysts. The technical term for *products* which can be bought or sold on the stock market is *Financial Instrument*. Financial instruments can be shares of a company (e.g., IBM shares). Financial instruments, however, can also be structured; that is, a financial instrument could relate to a fund that manages a portfolio of shares or even a hedge fund that manages a portfolio of certificates of other funds and hedge funds. It is in part this recursive nature of financial instruments that makes it difficult for business analysts to extract the right information from a data warehouse.

As mentioned in the introduction, real-world data warehouses are far more complex. In a real data warehouse, schemas are layered with different levels of abstraction. Figure 1 is a dramatic simplification, which could be at the *conceptual layer* at which business analysts and architects meet in order to design a new report. At lower layers (i.e., *logical* and *physical layer*), the schemas become more complex as the system architects and database administrators refine the schemas in order to achieve better performance by partitioning and replicating data, improve data quality by modeling data at different granularities, etc. Figure 2 shows the example schema at the logical level. Here, the addresses of individuals are actually stored in a separate table and transactions are modeled as either financial instrument transactions or money transactions.

It is easy to imagine how complex such schemas can get in a global financial institution considering varying regulatory requirements of different countries, redundancy that arises from keeping data in different granularities for performance reasons, the heterogeneity of the data sources that feed the data into the data warehouse, and with different departments asking for different kinds of reports.

## 2.2 Extended Metadata Graph

The data warehouse of Credit Suisse consists of base data stored in a relational database as well as metadata stored in a graph structure (such as RDF). The metadata consists of the database schema extended with DBpedia and domain ontologies (see Figure 3).
**Integrated Schema.** A data warehouse combines and aggregates data from many heterogeneous data sources. To handle the differences in the data sources, an integrated schema is built. To facilitate the design process, there are different levels of the schema, namely conceptual, logical and physical. The conceptual schema (business layer) serves for communication with business and contains the main entities to be modeled such as parties, transactions, and securities. The logical schema extends the conceptual one by showing inheritance, splitting entities (for instance, parties are split into individuals and organizations), etc. The physical schema contains information about database indexes or table partitioning. Typically all these schemas are designed with one modeling tool with the goal to generate the physical tables.
**Domain Ontologies.** In addition to the schema, our metadata consists of several domain ontologies. The domain ontologies are built for a given data warehouse and are used to classify data for a specific domain. As an example, such a domain ontology could classify financial instruments or customers. At Credit Suisse, customers are divided into private and corporate customers: Private customers are implemented using an *Individuals* table; the information of corporate customers is stored in an *Organizations* table.
**DBpedia.** The metadata also contains data from DBpedia in order to capture synonyms. Credit Suisse only maintains DBpedia entries that have direct connections to the terms stored in the integrated schema of the data warehouse. For instance, for the term "Parties" shown in our example world, the following entries have been extracted from DBpedia: *customer, client, political organization*, etc. As a result, when a user searches for *customers* then, *parties* would be one possible answer.
**Base Data.** As in most large-scale data warehouses, the base data of the Credit Suisse data warehouse is stored in relational databases. All the base data is implicitly connected to the metadata by the table and column names of the physical schema that holds the base data.

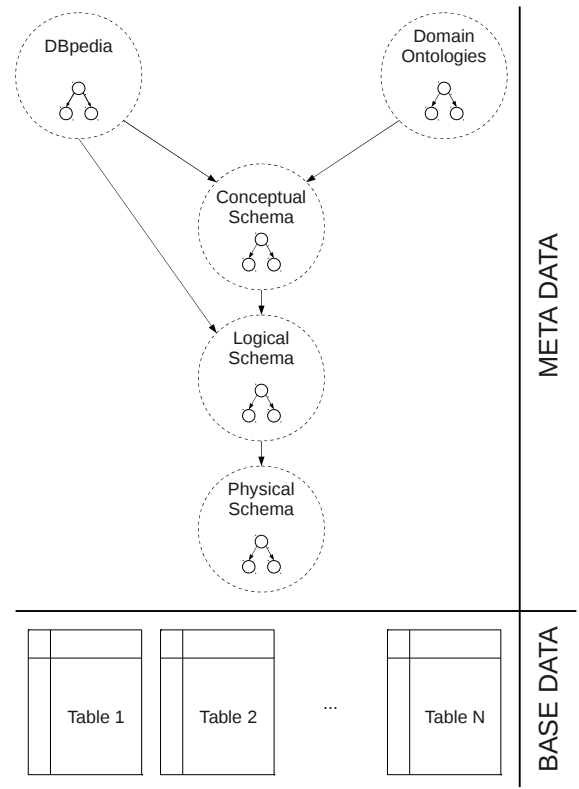

**Figure 3: Metadata Graph and Relational Data.**

## 3. SODA IN A NUTSHELL

Before elaborating on the patterns and the algorithms used in SODA, we would like to give a high-level overview of SODA [4]. Figure 4 shows the main steps of the SODA approach. These steps are similar to the way systems like BANKS, DBExplorer, and DISCOVER generate SQL queries. Again, the magic of SODA lies in the use of metadata and patterns (described in Section 4). Starting from a list of keywords and operators, SODA computes a ranked list of executable SQL statements that are likely to meet the information needs of the user. This transformation is carried out in the following five steps:

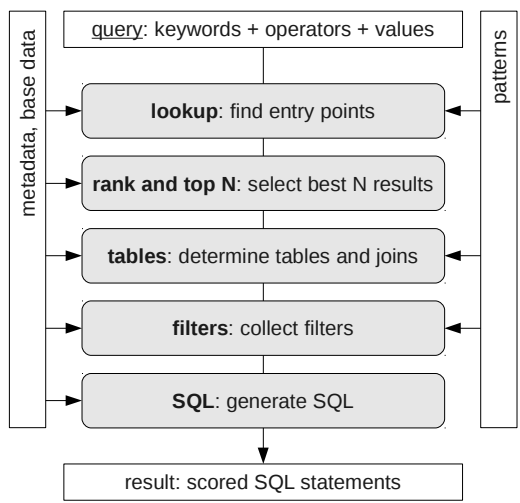

Figure 4: SODA Overview.

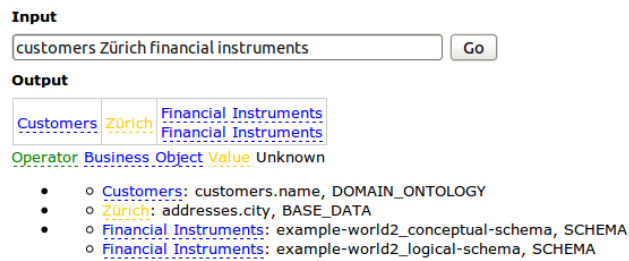

Figure 5: Query Classification.

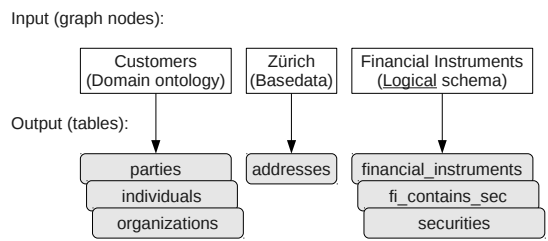

Figure 6: Output of Tables Step (join relationships not shown).

**Step 1 - Lookup:** The lookup step matches the keywords of the input query to sets of possible entry points. A lookup of a single keyword provides us with all the nodes in the metadata graph where this keyword is found. For example, in Figure 5, the keywords "customers" and "Zürich" are both found once, in the domain ontology and the base data, respectively. On the other hand, the keyword "financial instruments" is found twice: once in the conceptual schema and once in the logical schema. The output of the lookup step is a combinatorial product of all lookup terms. For this example two solutions are produced: One where "financial instruments" is found in the conceptual schema, and another one where "financial instruments" is in the logical schema. Besides processing keywords, our algorithm also uses operator constructs (patterns) to express aggregation and filters.

**Step 2 - Rank and top N:** The next step assigns a score to every result and continues with the best N results. For the ranking, we currently apply a simple heuristic which uses the location of the entry points in the metadata graph to assign a score to a result. For example, a keyword which was found in DBpedia gets a lower score than a keyword which was found in the domain ontology. We rank the domain ontology higher, because it was built by domain experts from the financial services industry, and hence it is more likely to match the intent of our business users than the general terms found in DBpedia. There exist certainly more sophisticated ranking algorithms such as BLINKS [9], however, ranking is only a part of SODA and not the main focus of this paper.

**Step 3 - Tables:** The purpose of this phase is to identify all the tables which are used in each solution and to discover the relationships between these tables. Starting at every entry point which we discovered in the lookup phase, we recursively follow all the outgoing edges in the metadata graph. At every node we test a set of graph patterns to find tables and joins. We assume, that tables found in this way, represent the entry points. In our example the output of this step are 7 tables (see Figure 6).

**Step 4 - Filters:** Filters can be found in two ways: a) by parsing the input query or b) by looking for filter conditions while traversing the metadata graph. In this step, we add the filters to the discovered tables and columns of the previous step. A filter condition consists of a column and a value such as "Zürich". In our example, the filter conditions are used to connect "Zürich" to the city column within the addresses table. While having filters in the input query is quite common, filters stored in the metadata can be very powerful as well. An example of a filter stored in the metadata would be "wealthy individuals" as described in the introduction.

**Step 5 - SQL:** In this final step, we take all the information that was collected earlier and combine it into reasonable, executable SQL statements. By "reasonable" statements we mean statements which take into account possible join patterns. For example, considering foreign keys and inheritance patterns in the schema. By "executable" statements we mean SQL statements that can be executed on the data warehouse.

## 4. GENERATING SQL FROM PATTERNS

This section provides details on the patterns and algorithms used in SODA. In particular, it shows how patterns are used to translate from a keyword-based input query to full-fledged SQL. Metadata graph patterns provide a flexible way to adapt the SODA algorithm to different data warehouses. It is important to note that patterns described in the following probably exist in all data warehouses, but the structure depends on the modeling of the data warehouse. Here, we describe the patterns that we used for Credit Suisse. To port SODA to a different data warehouse involves adjusting the patterns to the specific structures used in that data warehouse.

### 4.1 Why Patterns?

SODA uses patterns in two situations:
1) In *Step 1 - Lookup* (Figure 4): Instead of trying to use natural language processing to understand the input, we have a set of so-called *input patterns* that SODA understands. For example, every operator is a little pattern which combines an operation (comparison or aggregation) with values or business entities.
2) In *Step 3 - Tables and Step 4 - Filters* (Figure 4): When deriving the tables and join conditions from a given set of so-called *entry points* (graph nodes which represent words of the input query), SODA tests for so-called *metadata graph pattern* matches while traversing the metadata graph. A matching pattern tells us when we arrived at a special node which could be, for instance, a table, a foreign key or an attribute with a filter condition.

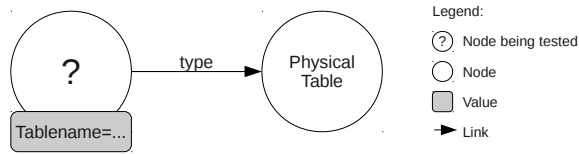

Figure 7: Table Pattern.

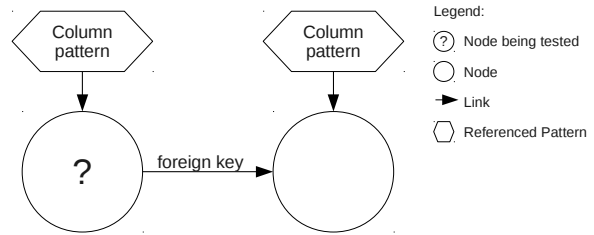

Figure 8: Foreign Key Pattern.

Both types of patterns, *input patterns* and *metadata graph patterns*, can be adapted to work for the given application. For example, use different input operators for another set of users or modify the metadata graph patterns according to the existing schema structure. While the patterns may have to be changed between different applications, the algorithm always stays the same.

## 4.2 Credit Suisse Patterns

As discussed previously, SODA works with two types of patterns: Input patterns and metadata graph patterns. Currently, input patterns are only keywords extended with a small set of operators, whereas our metadata graph patterns can match the complexity of the Credit Suisse metadata warehouse [11].

### 4.2.1 Metadata Graph Patterns

In *Step 3 - Tables and Step 4 - Filters* (see Figure 4), we use metadata graph patterns to discover tables, joins and filters stored in the metadata.

**Pattern Descriptions.** To define the patterns, SODA uses a language which was inspired by SPARQL [21] filter expressions: Each triple either connects two nodes or connects a node with a text label. A node is either a static URI or a variable. Variables can be assigned any URI, but within one match, a variable keeps its URI. An edge is a static URI. A text label is simply a string. In the following, we will use italic, dark gray font for variables, put t: before text labels, and remove URI prefixes for brevity.

To match a pattern on a given graph, we assign the variable $x$ to the current node and try to match each triple in the pattern to the graph accordingly.

**Basic Patterns.** These patterns describe how tables, columns, etc. are represented in the metadata graph. SODA matches these patterns against the metadata graph to identify the tables and columns which participate in each result. Basic patterns are used at the beginning of *Step 3 - Tables* of the SODA algorithm (see Figure 4).

The Table pattern can be written like this:

```
( x tablename t:y ) &
( x type physical_table )
```

This pattern matches, if the current node ($x$) has a `tablename` attribute pointing to a text label (`t:y`). In addition, $x$ needs to have a `type` attribute pointing to a node which has the URI `physical_table`. See Figure 7 for a graphical representation of these conditions. The Column pattern could be:

```
( x columnname t:y ) &
( x type physical_column ) &
( z column x )
```

The first part of the column pattern specification is similar to the table pattern. In the last line of the column pattern specification, we ensure that each column ($x$ in this example) has an incoming `column` edge from another node ($z$).

**More Complex Patterns.** These patterns define join relationships and inheritance structures. The simplest implementation of a join relationship is a direct edge between a foreign key attribute and a primary key attribute. This is shown in the Foreign Key pattern:

```
( x foreign_key y ) &
( x matches-column ) &
( y matches-column )
```

The term "matches-column" references the Column pattern described above. Figure 8 shows a visual representation of this pattern. In the case of Credit Suisse, we use a more general Join-Relationship pattern which has an explicit join node with outgoing edges to primary key and foreign key. For testing a node if it is a child in a inheritance structure, we use the Inheritance Child pattern:

```
( y inheritance_child x ) &
( y type inheritance_node ) &
( y inheritance_parent p ) &
( y inheritance_child c1 ) &
( y inheritance_child c2 )
```

Here, $x$ needs to have an incoming `inheritance_child` edge from an explicit inheritance node ($y$). The inheritance node, in turn, has to be of type `inheritance_node` and needs to have three outgoing edges: `inheritance_parent` to the inheritance parent, and two `inheritance_child` edges to the inheritance children.

**Application in SODA.** The metadata graph patterns described so far are all used in *Step 3 - Tables* of our algorithm (see Figure 4). We traverse the metadata graph starting from the entry points of a given query and recursively follow all outgoing edges. At each node, we test the Table, Column and Inheritance Child patterns. If the Table pattern matches, we collect the corresponding table name. If the Column pattern matches, we collect the corresponding column name as well as the table name. And if the Inheritance Child pattern matches, we collect the table name of the inheritance parent. We need to collect the table name of the inheritance parent because this table is needed to produce correct SQL statements.

After this first part of step 3 in the algorithm, we now have the table names for all given entry points. What remains to do in the second part of step 3, is to identify *joins* that are needed to properly connect the tables. Fortunately, a similar approach as for the table names can be used: We again traverse the metadata graph starting from the entry points of a given query and recursively follow all outgoing edges. But, instead of testing the Table, Column and Inheritance Child patterns as before, we now try to match the Foreign Key pattern (or, in the case of Credit Suisse, the Join-Relationship pattern). Of all the join conditions we discovered in this way, we now use these which are on a direct path between the entry points. Join conditions which are only "attached" to such a path are ignored to keep the result small and precise. A user interface, however, could make such joins available to the user. See Figure 9.

**Bridge Tables in Large Schemas.** Joining the entry points as described until now works fine in small data warehouses with simple schemas, i.e. in our example world, this works well. In a large data warehouse, as for example the Credit Suisse data warehouse, this is not enough. In a last part of step 3 of our algorithm, we therefore look for bridge tables, i.e. physical implementations of N-to-N relationships. Bridge tables connect two entities by having two outgoing foreign keys. If we find a bridge table between two of our entry points, we use it to add additional join conditions. Bridge tables—as you might have guessed—can be described with a pattern and identifying bridge tables therefore works similar to identifying tables and joins.

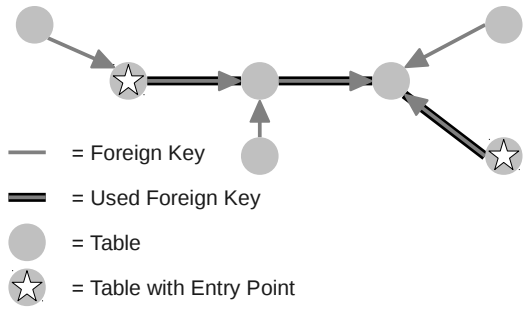

**Figure 9: Joins on Direct Path.**

### 4.2.2 Input Patterns

Input patterns are used in *Step 1 - Lookup* of the SODA algorithm (see Figure 4). These patterns are matched against the query terms to identify their meaning.

**Keywords.** The first type of input patterns are *keywords*. To process keyword-only inputs, we look for longest word combinations. We first try to match all the words in the input against our classification index. If we find a match, we are done. Otherwise, we recursively try smaller word combinations. In the following example, we find "Private customers" and "Switzerland":

```
Private customers Switzerland
```

Keyword-only inputs are the simplest way to use SODA and most people are familiar with using keywords for searching. The following types of input patterns are for more advanced users. Often, one starts with keywords only and afterwards adds operators to refine the query.

**Comparison Operators.** The second type of input patterns are *comparison operators*. Each comparison operator is a small binary pattern where the operator is in the middle and its operands are to the left and to the right. We currently support $>$, $>=$, $=$, $<=$, $<$ and *like*.

To identify operators and its operands in the input, we run our longest word combination algorithm as explained for Keywords above. This works well, because operators are simply words in the input and we can recognize them as such. The comparison operator will later on be applied to the keywords before and after itself.

**Aggregation Operators.** The last type of input patterns are *aggregation operators*. Here we currently use a very strict syntax, but this could be relaxed by modifying the pattern. We currently support $sum$ and $count$, however, there is nothing that would prevent us from adding more when we need it.

### 4.3 Query Language

Our query language for processing keywords and comparison operators can be formalized as follows:

```
<search keywords> [ [AND|OR]  <search keywords> |
<comparison operator> <search keyword> ]
```

The optional parts are written in parenthesis, i.e. between [ and ]. The pipe sign indicates "or". In order to express time-based range queries, the following syntax needs to be applied:

```
<search keywords> [ [AND | OR] <search keywords> |
<comparison operator> date(YYYY-MM-DD) ]
```

The characters Y, M, D refer to year, month and date.
The formal specification for aggregate queries is as follows:

```
<aggregation operator> (<aggregation attribute>)
[<search keywords>]
[group by (<group-by attribute1, ,attributeN>)]
```

Example queries for all types of input patterns can be found later in this section and in Section 5.

### 4.4 Patterns in Action - Examples

In this subsection we explain how we use patterns by looking at examples. We first present several queries which contain filter conditions. Afterwards we look at aggregation examples.

#### 4.4.1 Examples with Filters

Assume that an end-user wants to find all information about Sara Guttinger—a customer of the bank. The respective SODA query as well as the SQL query are shown in Query 1.

---
SODA:

   Sara Guttinger

SQL:

```
SELECT *
FROM parties, individuals
WHERE parties.id = individuals.id
      AND individuals.firstName = 'Sara'
      AND individuals.lastName = 'Guttinger'
```
---

**Query 1:** Keyword pattern example: SODA vs. SQL.

As we can see the SODA query is much easier to understand for a typical end-user than the SQL query where one needs to take into account the correct join and filter conditions.

In the second example we are looking for everyone who has a salary above a given value and was born on a certain date. The input query given to SODA is shown in the upper part of Query 2.

In this query we find three input patterns: a greater-equal comparison, an equality comparison, and a $date()$ operator. The remaining keywords are processed with our metadata graph pattern matching algorithm. Both, "salary" as well as "birthday", would match our column pattern and we would therefore include the corresponding table (e.g. persons) as well as the two attributes (persons.salary and persons.birthday). "and" might be unknown and we therefore ignore it. For completeness, a possible SQL query which does the same is also shown in Query 2.

#### 4.4.2 Examples with Aggregation

Assume that a business analyst wants to find the top n customers with the highest trading volume in 2010. Ideally, the input query would look like

```
Top 10 trading volume customer
between January 2010 and December 2010
```

```
SODA:
  salary >= x and birthday = date(1981-04-23)
SQL:
  SELECT *
  FROM persons
  WHERE persons.salary >= x
    AND persons.birthday = 1981-04-23
```
**Query 2:** Input pattern example: SODA vs. SQL.

Unfortunately, the first problem is that the given date range could refer either to trading volume or to customer. To resolve this ambiguity, one could write the query as a)

```
Top 10 trading volume customer transaction date
between date(2010-01-01) date(2010-12-31)
```

In this case, we would find customers with top trading volume with executed transactions in a certain time frame. On the other hand, if we wanted to find young costumers with high trading volumes, we would write the query as b)

```
Top 10 trading volume customer birth date
between date(1980-01-01) date(1990-01-01)
```

For this problem, it suffices to show both results to the business analyst and let her choose the result that matches her intent. The second problem, however, is more difficult to tackle: How to infer from "trading volume" to "aggregation of transaction amount"? One way to solve this problem is to let the user write a more precise query with explicit aggregation operator, e.g.

```
Top 10 sum(amount) customer transaction date
between date(1980-01-01) date(1990-01-01)
```

Since this is rather unintuitive, another way to handle such cases is to introduce a domain ontology. A domain ontology in our case defines a classification for a given domain.

To give a concrete example, we assume that we are interested in the amount of the transactions per trading day. The corresponding SQL statement and its SODA counterpart are shown in Query 3.

```
SODA:
  sum (amount) group by (transaction date)
SQL:
  SELECT sum(amount), transactiondate
  FROM fi_transactions
  GROUP BY transactiondate
```
**Query 3:** Aggregation pattern example: SODA vs. SQL.

The advantage of the SODA aggregation approach over standard SQL becomes even more apparent for aggregation queries that require a multi-table join. Since SODA automatically identifies the join predicates, the end-user does not need to worry about writing full SQL, which is often hard for typical non-tech savvy business analysts and hence SODA takes over that burden to enable user-friendly data warehouse search.

Consider, for instance, the example where we want to rank the organizations by trading volume. Query 4 shows the SODA query and the corresponding proper SQL statement. From the point of view of a business analyst, the SODA query is more intuitive, easier and much faster to write.

```
SODA:
  count (transactions) group by (company name)
SQL:
  SELECT count(fi_transactions.id), companyname
  FROM transactions,fi_transactions,organizations
  WHERE transactions.id = fi_transactions.id
    AND transactions.toParty = organizations.id
  GROUP BY organizations.companyname
  ORDER BY count(fi_transactions.id) desc
```
**Query 4:** Organizations ranked by trading volume example: SODA vs. SQL.

## 5. EXPERIMENTS & RESULTS

In this section we report on the experiments that we carried out with SODA on Credit Suisse's central data warehouse, which is among the largest and most complex data warehouses in the financial industry. Our experimental results demonstrate that SODA's keyword search algorithm generates executable SQL queries with high precision and recall compared to the manually written gold standard queries. (Gold standard queries have been manually written by domain experts.) SODA reveals ambiguities of the query keywords by understanding different patterns of the schema graph and by searching the base data using an inverted index. In addition to simple keyword search, SODA also supports conjunctive range queries as well as aggregations. As we will see in this section, the results clearly show that SODA not only works well for small schemas with several tables but also for large schemas with complex inheritance and join relationships of a modern enterprise data warehouse from the financial services industry. Moreover, we will highlight our experiences and challenges we faced when working with large and complex data sets.

### 5.1 Experimental Setup

#### 5.1.1 The Credit Suisse Data Warehouse

The Credit Suisse data warehouse is an enterprise data warehouse that consists of three main layers, namely *integration layer*, *enrichment layer*, and *analysis layer* [8]. The integration layer receives data from some 2,500 different source systems covering all areas of the bank such as information about customers, investment products, trades, etc. The total data volume covering three test environments and one production environment fully replicated over physically separated data centers is currently around 700 terabytes. The unintegrated data comprises several thousands of tables with some 30,000 attributes. The main goal of the integration layer is to take the data from the heterogeneous data sources and integrate them into a carefully modeled enterprise data warehouse with bi-temporal historization [5]. In other words, the data warehouse is a temporal database system with time dimensions covering the *validity time* and the *system time* [20].

Once the data is fully historized and quality controlled, the *enrichment layer* is used for storing so-called reusable measures and dimensions that are calculated based on previously integrated data. A typical example of data enrichment are SOX, Basel II, and Basel III calculations that compute complex base key figures that are materialized for efficiency reasons. The actual data analysis takes place in the *analysis layer* which consists of several business specific physical data marts fed either from the integration layer or the enrichment layer. Typical examples of these business applications are dedicated data marts for risk calculations, legal and compliance assessments or profitability calculations.

In addition to the actual data warehouse, Credit Suisse also has a *metadata warehouse* [11] that allows navigating and searching the complex schema of the various data warehouse layers. This metadata warehouse enables business users, requirements engineers, and software architects to get a better understanding of the complex relationships between the various data items. SODA builds on top of the metadata warehouse and provides additional functionality, namely automatic SQL generation and hence interactive data exploration based on keywords.

### 5.1.2 Software and Hardware Used

SODA is implemented in Java 1.6 with a generic database back-end that we tested with three different database systems: Derby, MySQL, and Oracle. Our experiments with real data were executed using Oracle 11gR1 on a Sun M5000 shared memory machine with 32 cores, 128 GB of main memory, and an enterprise-scale storage back-end that is attached to several data warehouse servers. The operating system is Solaris 10. Our data set is based on the full schema of the integration layer consisting of 472 tables with a reduced and anonymized data volume of 220 GB. The top 10% of these tables have above $10^7$ records with the largest table comprising 6.7 x $10^8$ rows. Moreover, the complex schema of the data warehouse consists of dozens of inheritance relationships with several levels.

Since our test environment is shared with other applications, our experiments were restricted to 4 cores and a maximum of 32 GB of main memory. Building up the inverted index for all 472 base data tables took 24 hours on a single core. The total size of the inverted index over all base tables is 9.5 GB comprising 1.4 x $10^8$ non-unique records. Note that the inverted index is only built on table columns of data type "text". In other words, base data table columns with numerical data types are not contained in our inverted index.

Table 1 shows the complexity of the schema graph in terms of the number of entities and attributes. Note that the cardinality of conceptual entities which represents the business world is 226, while the cardinality for logical entities and physical tables is 436 and 472, respectively. These numbers indicate that the complexity of the technical world increases with respect to the business world. Also note that the total number of attributes increases from 985 for the business world, that is, the conceptual attributes, to 2700 and 3181 for the technical world, that is, logical attributes and physical columns, respectively. The total size of schema graph is 37 MB which is relatively small compared to the total size of the base data which is 220 GB.

The great challenge that every business analyst faces who wants to query the data warehouse, is to understand the meaning of all these 436 entities and 472 tables and to correctly correlate them to each other (i.e. to understand the relationships among each other). As we can see in Section 5.3.2, SODA is considered as an important step to enable end-users to explore the data warehouse in an intuitive way.

### 5.1.3 Queries

Our query workload consists of a mix of queries taken from the query logs, queries proposed by our business users and synthetic queries to cover corner cases of our algorithms—such as complex aggregations with joins. Inspired by the 20 queries for astrophysics proposed by Jim Gray et al. [22], this paper shows the results for the 10+ most interesting queries we executed against one of the Credit Suisse data warehouses. These queries that are shown in Table 2 cover various query types such as queries against base data (B), against the schema (S) or domain onotolgy (D). Others handle inheritance (I), predicates (P) or aggregates (A). The abbreviations for the query types are shown in the column "Comments". Note that these different queries types are later on used for comparison with other systems (see Table 5)).

**Table 1: Complexity of the schema graph including conceptual, logical and physical schema.**

| Type | Cardinality |
|---:|---:|
| #Conceptual entities | 226 |
| #Conceptual attributes | 985 |
| #Conceptual relationships | 243 |
| #Logical entities | 436 |
| #Logical attributes | 2700 |
| #Logical relationships | 254 |
| #Physical tables | 472 |
| #Physical columns | 3181 |

Column 2 shows the queries expressed in terms of keywords and column 3 gives additional comments about the queries. For instance, query Q1.0 is answered by finding a match of "private customers" in the customer domain ontology as well as finding a match of "family name" in the schema graph. Query Q2.1 is evaluated by using "Sara" as a filter criterion on the base data. Queries Q2.2 and Q2.3 are additional refinements of query Q2.1 to yield more precise answers. Queries Q3.1 and Q3.2 show the ambiguity of the query "Credit Suisse". The user could either be interested in Credit Suisse as an organization or as an entity that is part of an agreement. Query Q5.0 is interesting since it needs to correctly identify inheritance relationships (by applying the inheritance pattern), while queries Q9.0 and Q10.0 are aggregation queries.

For each of these queries we manually wrote so-called *gold standard* queries (see column 4), i.e. executable SQL statements that return the best results for the given queries. The gold standard queries serve as the yard stick for measuring precision and recall of the SODA queries.

## 5.2 Results

### 5.2.1 Precision and Recall

For each input query $Q_i$, we compute precision and recall for all $j$ SQL statements $R_{ij}$ SODA produces. To compute precision, we compared the result tuples of a produced SQL statement of SODA $\#R_{ij}$ with the result tuples of the Gold Standard query $\#G_i$. A precision of 1.0 means, that a SQL statement produced by SODA returned only tuples that also appear in the Gold Standard result $\#R_{ij} \subseteq \#G_i$. Similarly, a recall of 1.0 means, that a SQL statement produced by SODA returned all tuples of the Gold Standard result $\#G_i \subseteq \#R_{ij}$. In Table 3 we show precision and recall of the best SQL statement produced by SODA. We also calculated the number of SODA results with precision and recall greater than 0 and equal to 0. Table 3 shows the results.

We can see that for a majority of the queries, SODA produces a precision of 1.0 while the recall is either 1.0 or 0.2 (see Q2.1 and Q2.2). The reason for the sometimes low recall is due to the fact that the data warehouse uses bi-temporal historization where the actual join keys are not properly reflected in the schema graph. In order to mitigate this problem, the schema graph needs to be annotated with join relationships that reflect bi-temporal historization. Note that SODA provides a very flexible way of incorporating these changes that are typically required for modern data warehouses that constantly evolve over time—both in size and complexity.

Another interesting observation is that for some queries the precision is close to 0 or even 0 (see Q5.0 and Q9.0). These results are

939

Table 2: Experiment queries.

| Q | Keyword | Comment | Gold standard |
|---|---------|---------|---------------|
| 1.0 | private customers family name | Use customer domain ontology (D) and combine with attribute from schema (S). | 3-way join incl. inheritance (I). |
| 2.1 | Sara | Use base data (B) as a filter criterion. | 3-way join incl. inheritance (I) with where-clause on given name. |
| 2.2 | Sara given name | Same as for Q 2.1 + restriction on given name (S). | Same as for Q 2.1. |
| 2.3 | Sara birth date | Restriction on birth date to focus on specific table (S). | Same as for Q 2.1. |
| 3.1 | Credit Suisse | Use base data (B) as a filter criterion to find the organization. | Single table containing information about organizations with where-clause on org_name. |
| 3.2 | Credit Suisse | Use base data (B) as a filter criterion to find Credit Suisse agreements. | Single table containing information about deals with where-clause on agreement_name. |
| 4.0 | gold agreement | Use base data (B) as filter and match with schema attribute (S). | 2-way join. |
| 5.0 | customers names | Identify inheritance relationships (I) and use names domain ontology (D). | Two separate 3-way join queries for private and corporate clients. |
| 6.0 | trade order period > date(2011-09-01) | Time-base range query (P) on given column (S). | 3-way join with where-clause incl. inheritance (I). |
| 7.0 | YEN trade order | Use base data (B) filters and schema (S). | 5-way join with 2 where-clauses incl. inheritance (I). |
| 8.0 | trade order investment product Lehman XYZ | Base data (B) + schema (S). | 5-way join with where-clause incl. inheritance (I). |
| 9.0 | select count() private customers Switzerland | Base data (B)+ domain ontology (D) + aggregation (A) | 5-way join + aggregation incl. inheritance (I). |
| 10.0 | sum(investments) group by (currency) | Aggregation (A) with explicit grouping and schema (S). | 5-way join + aggregation + group by. |

Table 3: Precision and recall for experiment queries including inverted index for base data.

| Q | Best Result Precision (P) | Recall (R) | #Results P,R > 0 | #Results P,R = 0 |
|---|---|---|---|---|
| 1.0 | 1.00 | 1.00 | 1 | 0 |
| 2.1 | 1.00 | 0.20 | 1 | 3 |
| 2.2 | 1.00 | 0.20 | 1 | 1 |
| 2.3 | 1.00 | 1.00 | 1 | 2 |
| 3.1 | 1.00 | 1.00 | 2 | 4 |
| 3.2 | 1.00 | 1.00 | 3 | 3 |
| 4.0 | 1.00 | 1.00 | 1 | 3 |
| 5.0 | 0.12 | 0.56 | 1 | 4 |
| 6.0 | 1.00 | 1.00 | 2 | 0 |
| 7.0 | 0.50 | 1.00 | 1 | 3 |
| 8.0 | 1.00 | 1.00 | 2 | 2 |
| 9.0 | 0.00 | 0.00 | 0 | 6 |
| 10.0 | 1.00 | 1.00 | 1 | 5 |

due to the complex nature of the data model with several bridge tables where SODA is not able to identify the correct join conditions.

### 5.2.2 Query Complexity and Runtime

After measuring the precision and recall of the queries, we now analyze the query complexity and the runtime.

The query complexity is defined as the number of combinations that can potentially lead to a query result. For instance, recall the query "customers Zurich financial instruments" shown in Figure 5. This query has a complexity of 1 x 1 x 2 = 2 which is explained as follows: The term "customers" occurs 1 time in the domain ontology, the term "Zurich" occurs 1 time in the base data and the term "financial instruments" occurs 2 times (once in the conceptual and once in the logical schema). In general, the number of results after the lookup phase grows quickly due to the combinatorial product of all entry points. The remaining steps, however, are all linear in the size of the meta-data.

The end-to-end execution time of a SODA query is split up into time fractions that correspond to the algorithmic steps: (1) lookup, (2) rank, (3) tables, (4) SQL and (5) grouping. The total time to execute these 10+ queries was roughly one hour where the majority of the time was spent on executing the generated SQL queries. The time for SODA to analyze the query and to produce proper SQL is in the range of seconds. Detailed numbers are given in Table 4. We can see that the SODA runtimes are between 0.73 and 7.31 seconds while the total runtime for executing the SQL query on the database ranges between 1 and 40 minutes. Note that query Q10.0 has the largest total runtime of 40 minutes due to the aggregation and group by operations that need to be performed along with a 5-way join on large tables. These numbers indicate that the overhead for the SODA query processing is a small fraction compared to the total query execution time.

## 5.3 War Stories

### 5.3.1 Experience and Challenges

In this section we discuss our experience and challenges we faced when working with the enterprise data warehouse of Credit Suisse. Perhaps one of the biggest challenges in a real data warehouse is *data quality*. On the one hand, the number of source systems and the constantly changing business requirements make it almost impossible to have a perfectly matching schema description that 100% reflects the physical database implementation. Moreover, data warehouses are actually never really finished, since new feeder files are ingested into the data warehouse on a regular basis. These, in turn, need to be modeled and integrated into the existing enterprise data model.

**Table 4: Query complexity and runtime information of SODA algorithm (sec) and total end-to-end query processing (min).**

| Q | Complexity | #Results | SODA runtime (sec) | Total runtime (min) |
|---|---|---|---|---|
| 1.0 | 3 | 1 | 1.54 | 6 |
| 2.1 | 4 | 4 | 0.81 | 1 |
| 2.2 | 12 | 2 | 1.60 | 3 |
| 2.3 | 12 | 3 | 1.69 | 3 |
| 3.1 | 12 | 6 | 3.78 | 2 |
| 3.2 | 12 | 6 | 3.78 | 2 |
| 4.0 | 16 | 4 | 4.89 | 4 |
| 5.0 | 4 | 4 | 1.24 | 6 |
| 6.0 | 5 | 2 | 0.73 | 1 |
| 7.0 | 20 | 4 | 4.94 | 1 |
| 8.0 | 8 | 4 | 2.94 | 2 |
| 9.0 | 30 | 6 | 7.31 | 1 |
| 10.0 | 25 | 6 | 2.83 | 40 |

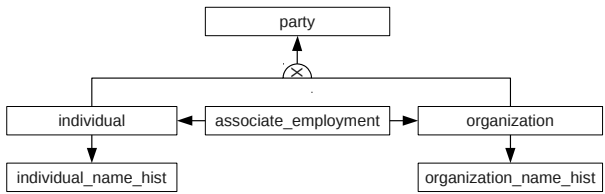

**Figure 10: Complex Schema Hierarchy**

Another real challenge is the *complexity of the schema* which includes inheritance relationships of several levels. The complexity increases when there are several relations (physical bridge tables) between siblings of an inheritance relationship (see entity "associate_employment" Figure 10). These bridge tables between siblings are common in several areas of the schema. Hence, automatic generation of SQL that takes into account these complexities is non-trivial—especially when some of the primary/foreign key relationships are not always implemented or the data is not populated yet. These issues often lead to queries with low precision and recall as we have seen for query Q5.0.

The strength of SODA is that these data quality issues or schema complexities can be mitigated by annotating the schema graph or by extending the graph patterns. For instance, if we know from—let's say the Testing Team—that some database tables that are part of a bridge tables between siblings are not populated yet, the schema can be annotated indicating that the respective relationship should be ignored. Once the underlying database tables are populated, the annotation can be updated so that SODA can take this relation into account when generating queries.

Another open issue of SODA is how to deal with temporal aspects of the data warehouse, i.e. bi-temporal historization. At the moment, SODA has no special support for temporal operators. Time is processed by SODA just like any other dimension. That is, SODA can generate queries that involve time (via range predicates on "valid-date" in a bi-temporal database or by restricting a year or a quarter), but SODA does not support, say, temporal aggregations or history joins. We plan to implement those as part of future work if business users ask for it.

Finally, SODA does not blindly produce all theoretically possible join paths, but rather combines a directed graph traversal with a given set of patterns to find useful tables and joins. While this has the advantage of being less computationally intensive and usually still leads to the intended results, there is no guarantee that we are not missing a required join path. E.g. we might not be able to find a join path between two entities which are too far apart in the schema graph. In this situation, "far-fetching" patterns might help. In other situations, however, "far-fetching" patterns might produce so many results that even ranking them becomes infeasible.

### 5.3.2 Feedback from Various Audiences

We demonstrated SODA to various people inside and outside of Credit Suisse to get feedback about our system. The people were both computer scientists as well as business users. One of the interesting observations is that different users see potentially completely different usage scenarios for SODA.

One group of people is impressed by the feature of the inverted index on the base data which allows identifying data items spread across several tables in the data warehouse that they were not even aware of. The reason for the data items to be located in different tables is due to the different data semantics.

Another group of people sees the potential of using SODA as an exploratory tool to analyze the schema and learn patterns in the schema in order to find out which entities are related with others. These types of users would issue a query and get a table as a result. Next, they would use the SODA schema browser to dive deeper. By an interactive approach of generating automatic queries based on keywords and analyzing the schema, they would identify potential flaws in the schema design or data quality issues.

A third group of users would use SODA to help creating SQL statements. They appreciate the feature that SODA automatically discovers join relationships between tables. Typically, end-users would just say "Give me tables X, Y and Z and show me the differences in calculations with respect to the previous months". These types of business users are not willing to define the complex join conditions by themselves. In fact, within Credit Suisse we incorporated some of the SODA functionality in the so-called Adjustment-Engine—a system that enables business users to adjust data in the data warehouse by themselves.

Finally, a forth group of people is looking into the possibility of using SODA as a way to help document legacy systems by reverse engineering the conceptual, logical and physical schema based on the existing physical implementation of the data warehouse. After the reverse engineering is completed, the RDF schema graph can be generated and annotated accordingly. SODA would give them the possibility to explore legacy systems where documentation is very scarce or does not even exist.

As we can see from the various types of feedback illustrated above, SODA can be used for different tasks that were originally not even foreseen when we designed SODA.

## 6. RELATED WORK

### 6.1 Search in Relational Databases

The design of SODA is based on the experience gained with a number of related systems that were developed over the last decade. The first systems to support keyword search in relational databases were DBExplorer [1], DISCOVER [10], and BANKS [3]. The key idea of these systems was to build an inverted index on the base data and to consider key/foreign key relationships when building query results. The inverted index is used to find all occurrences of the keywords of a query in tuples of the database. The key/foreign key relationships are used to compute join paths to construct business objects from the tuples that match different keywords of the query. The results of DISCOVER and BANKS are in the granularity of specific instances (i.e., individual business objects assembled from individual tuples that match the keywords). DBExplorer generates results in the granularity of sets of business objects. All three approaches differ in the way they generate the join paths.

Based on the foundations laid with this early work on keyword search in relational databases, a number of more sophisticated systems have been developed in the recent past. Keymantic [2] shows how to support search on the "Hidden Web". In the "Hidden Web", no inverted indexes can be constructed because the base data is not crawlable. The only information that is known to Keymantic is metadata such as the names of input fields from, e.g., crawling the Web forms of a "Hidden Web" database. So, a keyword query is processed as follows: First, all keywords that correspond to metadata items (e.g., field names) are extracted. The remaining keywords are considered as possible input fields. Second, the likelihood of a remaining keyword to a metadata item is computed in order to rank different options to execute the keyword query on the "Hidden Web" database.

The work that is most closely related to SODA is SQAK [23]. SQAK is the only system that we are aware of that is able to generate aggregate queries. It is, therefore, well suited for data warehouses. SQAK has, furthermore, a special way to compute join paths that respects the direction of key/foreign key relationships. Unfortunately, all these techniques are hard-coded into the SQAK approach. As a result, SQAK is not able to process any queries that go beyond the pre-defined SQAK pattern of *SELECT-PROJECT-JOIN-GROUP-BY* queries. Furthermore, SQAK is not able to integrate metadata in the flexible and general way that SODA can.

### 6.2 Evaluation of Related Systems

Table 5 gives an overview of which features are supported by the related systems described in the previous sub-section. It shows the features that are supported by the individual systems and which benchmark queries involve these features. Keymantic was the only system that we could evaluate experimentally because executable binaries were available from the authors; for the other systems, the overview of Table 5 is based on the description from the papers.

Simple queries that involve keywords found in the base data (e.g., "Sara" or "Credit Suisse") are obviously well supported by DBExplorer, DISCOVER, and BANKS, as shown in the first line of Table 5. Since SQAK specifically targets aggregate queries, it cannot handle simple keyword queries; such simple *SELECT* queries just do not match SQAK's predefined pattern. In principle, Keymantic can handle such simple keyword queries, but for complex schemas with thousands of columns like that of the Credit Suisse data warehouse, Keymantic is not able to select the right columns to query even when given all the available metadata. It should be noted that DBExplorer as well as DISCOVER cannot handle even simple queries if the schema involves cycles. So, these two systems sometimes have issues for simple keyword queries on base data (indicated by a check mark in parenthesis in Table 5).

The advantages of SODA only become apparent for more complex queries and for queries that mix several features and involve metadata. While DBExplorer, DISCOVER, and BANKS do support look up of keywords in base data, these systems are nevertheless not able to process, e.g., Query 9 because that query also involves the right treatment of inheritance, domain ontologies, and group-by / aggregation. As a result, each of the systems listed in Table 5 (except SODA) can handle only a few of the benchmark queries (and those with caveats).

The only other system that is able to integrate metadata beyond key/foreign key relationships is Keymantic. To some extent it can handle queries that involve synonyms and homonyms (i.e., queries that involve a domain ontology or DBpedia data). But, even Keymantic cannot handle any queries that involve inheritance. Modeling inheritance involves the modeling of mutually exclusive relationships. Even within Credit Suisse such inheritance relationships are not modeled in a consistent way; that is why a flexible pattern matching approach is needed as used in SODA in which different patterns can be specified for the same concept. Flexible pattern matching is even more important in a generic search tool that is supposed to be used in different organization with highly varying modeling conventions.

It is worth mentioning that in data warehouses such as those found at Credit Suisse, physical column and table names never correspond to those documented as part of a conceptual or logical schema. At Credit Suisse, for example, "birth date" is shortened to "birth dt"; furthermore, entity names (such as agreement or investments) are suffixed with "td". The best way to discover such matches is to keep metadata at multiple schema levels and to apply pattern matching across those levels as done in SODA (Figure 3).

SODA is also the only system that can properly deal with predicates. While it is conceivable that some of the systems be extended to deal with certain kinds of range predicates (e.g., date ranges), SODA is the only system that is able to handle predicates induced by the metadata (e.g., *wealthy customers* as customers that have an annual income that is higher than a certain threshold defined as part of the domain ontology or other metadata).

### 6.3 Other Related Work

The systems discussed in the previous two sub-sections are not the only related work. Various aspects of generating SQL from keywords have been studied in the literature. For instance, [9] studies alternative ranking algorithms; [15] addresses physical database optimization by using more efficient index structures; [19] supports complex queries by a more sophisticated approach to process natural language; and [18] provides tuple reduction. Other works use principles from information theory and statistics to summarize the relational schemas [24].

Another line of research studies the use of query refinement and query disambiguation approaches [17, 6, 7]. Ortega-Binderberger et al. [17] studies the importance of user subjectivity and achieves query refinement through relevance feedback. Similarly, SODA presents several possible solutions to its users and allows them to like (or dislike) each result. Elena Demidova et al. [6, 7] use query disambiguation techniques to process keyword queries automatically extracted from documents.

SnipSuggest [12] is a system that enables context-aware auto completion for SQL by taking into account previous query workloads which are, in turn, represented as workload DAG. When a user types a query, possible additions are suggested based on the highest ranked node in the DAG. Query suggestions include tables, views, functions in the FROM-clause, columns in the SELECT and GROUP BY clauses as well as predicates in the WHERE clause. The main difference to our approach is that SnipSuggest makes it easier for end-users to interactively build and improve SQL statements while SODA does not require any SQL knowledge at all. Moreover, SODA does also not rely on query logs.

Keyword search [16] and natural language processing [14] have also been applied to XML databases. [16] presents a survey that

Table 5: Qualitative comparison.

| Query type | Experiment Queries | DBExplorer | DISCOVER | BANKS | SQAK | Keymantic | SODA |
|---|---|---|---|---|---|---|---|
| Base data | 2.*, 3.*, 4, 7, 8, 9 | (✓) | (✓) | ✓ | NO | (NO) | ✓ |
| Schema | 1, 2.2, 2.3, 4, 6, 7, 8, 10 | NO | NO | ✓ | NO | ✓ | ✓ |
| Inheritance | 1, 2.*, 5, 6, 7, 8, 9 | NO | NO | NO | NO | NO | ✓ |
| Domain ontology | 1, 5, 9 | NO | NO | NO | NO | (✓) | ✓ |
| Predicates | 6 | NO | NO | NO | NO | NO | ✓ |
| Aggregates | 9, 10 | NO | NO | NO | ✓ | NO | ✓ |

classifies search methods into four categories: a) Tree-based methods, where the result is based on the notion of LCA (lowest common ancestor). b) Statistics-based approaches, which work with statistics on the data distribution. c) Graph-based methods, which look for connecting subgraphs containing all keywords. d) Methods on RDF graphs, where the additional semantics of an RDF graph are utilized. SODA is closely related to methods that fall into categories (c) and (d). While the presented approaches can work with RDF graphs, they are not really making use of the additional semantics. The patterns in SODA, however, allow us to capture this information. The NaLIX system [14] takes natural language query as input and translates it into XQuery. One of the strengths of the system is that it provides feedback to the user if the query terms cannot be classified and hence translated. In these cases, the system suggests different ways of reformulating the queries.

## 7. CONCLUSIONS

In this paper we demonstrated that SODA (Search Over DAta Warehouse) is one step towards enabling end-users to interactively explore large data warehouses with complex schemas in a Google-like fashion. The key idea of SODA is to use a graph pattern matching algorithm to generate SQL based on simple key words. Our experiments—with both synthetic data as well as with a large data warehouse of a global player in the financial services industry—show that the generated queries have high precision and recall compared to the manually written gold standard queries. One of the strengths of SODA is that it can disambiguate the meaning of words by taking into account join and inheritance relationships among the matching tables. Moreover, SODA allows mitigating inconsistencies in the schema or data as well as data quality issues by updating the respective metadata graph or by extending the graph pattern match algorithm.

As part of our future work we will evaluate the impacts of using DBpedia for matching keyword queries against various synonyms found in our classification. Since the use of DBpedia will naturally increase the number of possible query results—the query complexity, we will study more advanced ranking algorithms. Furthermore, the current GUI of SODA could be extended in several ways to engage the user in selecting and ranking the different results. Finally, we plan to use additional metadata graph patterns, for example, to better cope with bi-temporal historization or data lineage across different layers of the Credit Suisse data warehouses.